\documentclass[twocolumn,english,prl]{revtex4-1}
\usepackage{color}
\usepackage{graphicx}
\usepackage[T1]{fontenc}
\usepackage{float}
\usepackage{textcomp}
\usepackage{amsmath}
\usepackage{amssymb}
\usepackage{graphicx}
\usepackage{esint}
\usepackage{natbib}
\usepackage{color}

\begin{document}

\title{Imaging  backscattering in graphene quantum point contacts}

\author{A. Mre\'{n}ca-Kolasi\'{n}ska}
\affiliation{AGH University of Science and Technology, Faculty of Physics and
Applied Computer Science,\\
 al. Mickiewicza 30, 30-059 Krak\'ow, Poland}

\author{B. Szafran}
\affiliation{AGH University of Science and Technology, Faculty of Physics and
Applied Computer Science,\\
 al. Mickiewicza 30, 30-059 Krak\'ow, Poland}
\begin{abstract}
We study graphene quantum point contacts (QPC) and imaging of the  backscattering of the Fermi level wave function by potential introduced by  a scanning probe.
We consider both etched single-layer QPCs as well as the ones formed by bilayer patches deposited at the sides of the monolayer conducting channel using an atomistic
tight binding approach.
A computational method is developed to effectively simulate an infinite graphene plane outside the QPC using a computational box of a finite size. 
We demonstrate that in spite of the Klein phenomenon interference due to the backscattering at a circular n-p junction induced by the probe potential is visible  in spatial maps of conductance
as functions of the probe position. 
\end{abstract}

\maketitle

\section{Introduction}
Quantum point contacts \cite{Beenakker,Wees} (QPC) are  elementary building blocks of quantum transport devices for carrier injection and read-out with 
control over the quantized conductance. 
Transport phenomena for the current injected through QPCs are studied with the spatial resolution by the scanning gate microscopy (SGM) \cite{SellierRev} --  a technique in which a charged tip of the atomic force microscope  perturbs the potential within the system with the 2DEG,  induces the backscattering and alters the conductance. 
 SGM was used for 
graphene-based systems, the QPCs \cite{Neubeck}
 states localized within the QPC  \cite{Connolly, Pascher, Garcia},  quantum Hall conditions \cite{Connolly2012, Rajkumar, Chua}, and
 magnetic focused  trajectories \cite{MF4, Bhandari}.
Theoretical studies for the magnetic focusing \cite{Petrovic} and imaging snake states \cite{SnakiKolacha} were 
 performed.

SGM for QPCs defined within the two-dimensional electron gas for heterostructures based on III-V semiconductors resolves interference of the incident and backscattered  \cite{branche0,branche1,branche2,branche3,branche4} wave functions.
In graphene, a strong tip potential induces formation of a local n-p junction \cite{TipKlein1} instead of depletion of the electron gas as in III-V semiconductors.
The n-p junctions in graphene are transparent for Fermi level electrons incident normally
due to the Klein tunneling \cite{Katsnelson2006, Allain2011,Cayssol,Schulz}. Nevertheless, as we show below, the backscattering induced
by the n-p junction formed by the tip  induces a clear interference in SGM maps
 with a period of half the Fermi wavelength.

\begin{figure}
 \includegraphics[width=0.7\columnwidth]{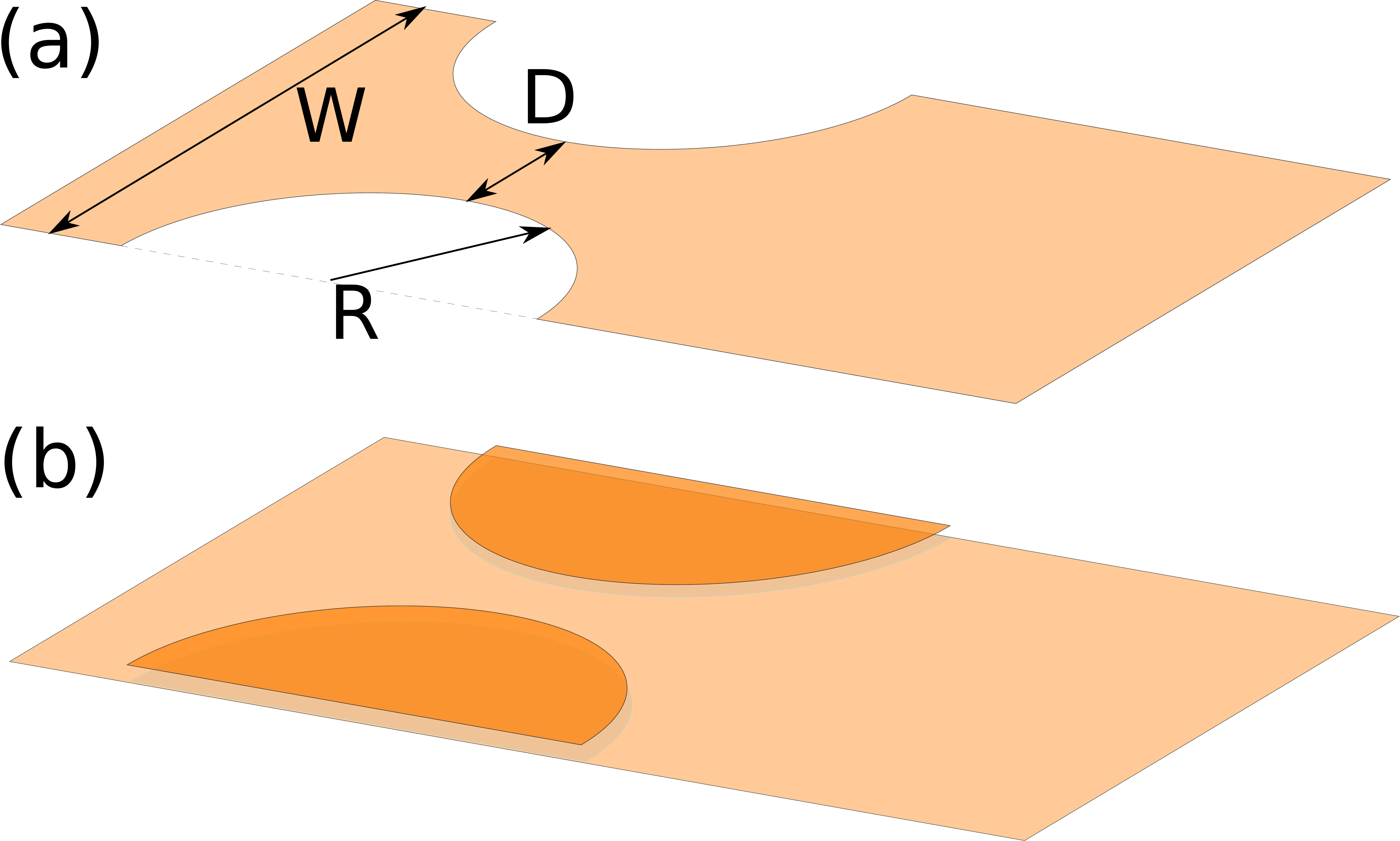}
 \includegraphics[width=0.86\columnwidth]{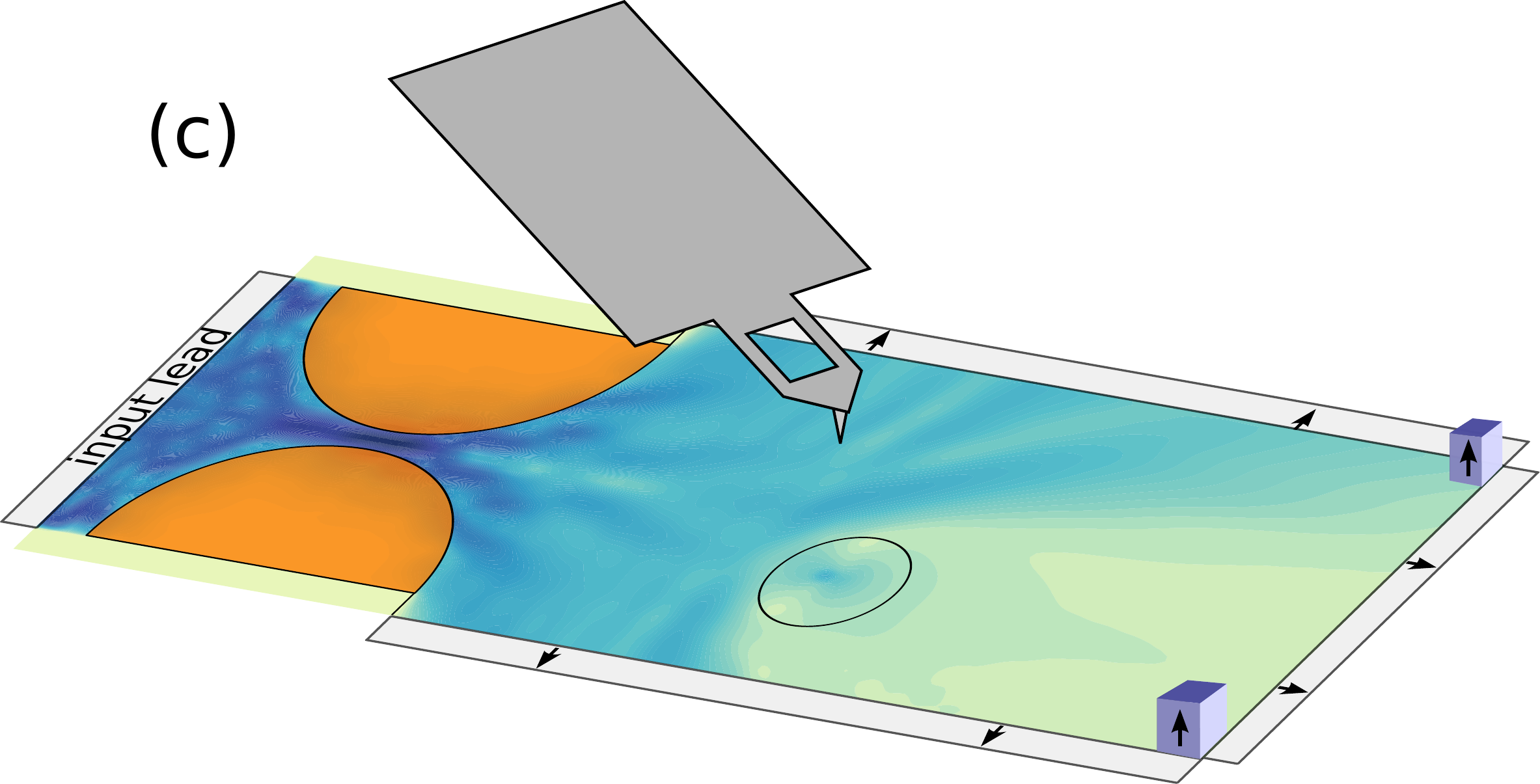}
  \caption{QPC etched out of graphene (a) and (b) built of patches of bilayer graphene. (c) Schematic drawing of the simulated scanning gate microscopy. The circle indicates the n-p junction for the tip potential equal to the  $V=E_F$. The light grey lines with the arrows indicate
the open BCs introduced as leads. The electric blue blocks
with the vertical arrows mark the additional leads used as a sink of currents to suppress backscattering by the corners.
  } \label{etch_ind}
\end{figure}

In semiconductor heterostructures with two-dimensional electron gas (2DEG), QPCs can be defined 
by lateral gates, which deplete the 2DEG, and thus change the constriction width and 
narrow the conduction channel for Fermi level electrons
 \cite{Wees}. 
In graphene the channel constriction by 
external gates is ineffective due to Klein tunneling \cite{yang2011}.
Etched QPCs were studied instead by both experiment \cite{Lin,Tombros2011,Terres, Kinikar} and theory \cite{Munoz,Ihnatsenka,IhnatsenkaHartree}.
In bilayer graphene \cite{McCann,McCann2,Castro} it is possible to induce a bandgap by applying a bias between the layers \cite{Neto,Xia,Park}. QPCs on graphene with bilayer inclusions have been produced \cite{Chua}, but conductance quantization in these systems has not been theoretically investigated so far.
For demonstration that the present results are independent of the QPC
type 
we consider both etched [Fig.\ref{etch_ind}(a)] and bilayer patched QPCs [Fig.\ref{etch_ind}(b)].
The latter 
turn out less susceptible to perturbation by defects within the QPC.

In order to discuss the effects of the sample imperfections 
we consider both defects at the edge and within the bulk of the sample.
For the edge deffects  we consider singly-connected carbon atoms \cite{He} protruding from the zigzag segments of the constriction edge that produce resonant scattering that destabilizes the conductance plateaux.
For the bulk imperfections we consider local potential perturbation introduced by fluorine adatoms deposited on the surface \cite{Irmer2015}.

%
%

\section{Theory}
We use the atomistic tight-binding Hamiltionian spanned by $p_z$ orbitals,
\begin{equation}
   H=\sum_{\langle i,j\rangle }\left(t_{ij} c_i^\dagger c_j+h.c.\right)+\sum_i V({\bf r}_i) c_i^\dagger c_i, 
\label{eq:dh}
\end{equation}
where  $V({\bf r}_i)$ is the external potential at the $i$-th site at position $\mathbf{r}_i$, and in the first term we sum over the nearest neighbors. We use the tight-binding parametrization of Ref.~\onlinecite{Partoens}, with 
$t_{ij}=-3.12$ eV for the nearest neighbors within the same layer.  For the bilayer, we take $t_{ij}=-0.377$ eV for the A-B dimers, $t_{ij}=-0.29$ eV for skew interlayer hoppings \cite{Partoens}  between atoms of the same sublattice (A-A or B-B type), and $t_{ij}=0.12$ eV for skew interlayer hopping between atoms of different sublattices. 
The potential energy in the lower layer is taken as the reference level $V_{b}'=0$, and  the value of the upper layer $V_{b}$ is tuned by the electric field perpendicular to the layer. The interlayer distance  is 3.32 \AA. 

In order to account for the effects of the lattice imperfections far from the edges of the sample we consider separate fluorine adatoms
with the tight-binding parametrization of the  hopping parameters taken from Ref.~\onlinecite{Irmer2015} in the dilute fluorination limit.
Accordingly \cite{Irmer2015}, for the hopping between the fluorine atom and the carbon ion 
we take $T=5.5$ eV and the on-site energy on the fluorine ion is $\varepsilon_F=-2.2$ eV. 

For simulation of the SGM, we assume an effective potential of the tip with a Lorentzian form \cite{kolasinskiDFT2013}
\begin{equation} V(x,y)=\frac{V_t}{1+\left( (x-x_t)^2+(y-y_t)^2\right)/d^2}, \label{lf} \end{equation}
where $x_t,y_t$ are the tip coordinates, $d$ is the effective width of the tip potential, and $V_t$ is its maximal value ($V_t=1.25$ eV  unless stated otherwise).

We consider the energy range near the Dirac point. 
For evaluation of the transmission probability, we use the wave function matching (WFM) technique as described in Ref. \onlinecite{Kolacha}. The transmission probability from the input lead to mode $m$ in the output lead is
\begin{equation}
T^m = \sum_{ n } |t^{mn}|^2,
\end{equation}
where $t^{mn}$ is the probability amplitude for the transmission from the mode $n$ in the input lead to mode $m$ in the output lead. We evaluate the conductance as $G={G_0}\sum_{m} T^{m}$, with $G_0={2e^2}/{h}$.

\begin{equation}
G= G_0 \sum_{ m,n } |t^{mn}|^2.
\end{equation}

\begin{figure}
 \hspace{0.2cm}   \includegraphics[width=0.82\columnwidth]{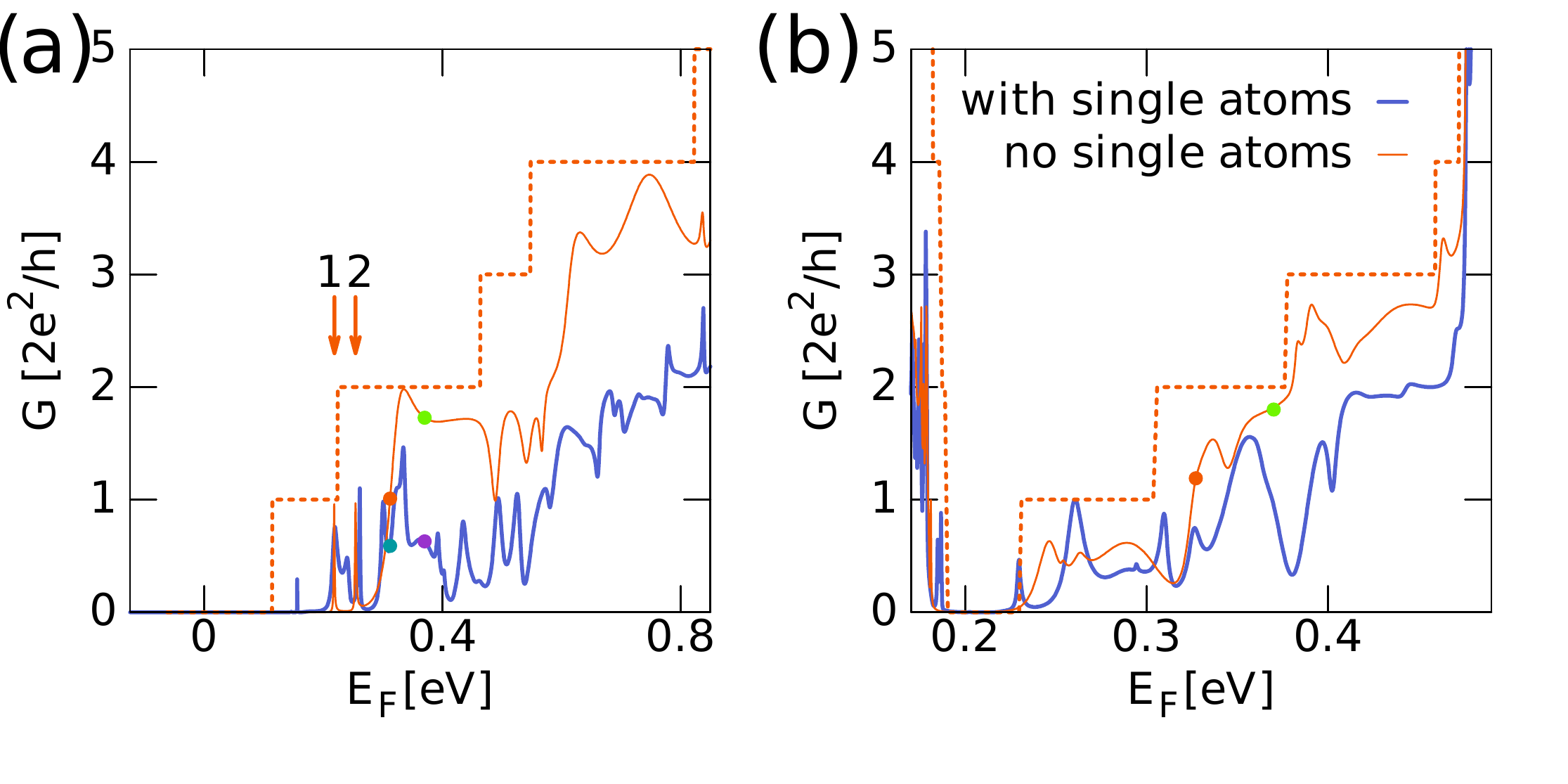}\\
\includegraphics[width=0.35\columnwidth]{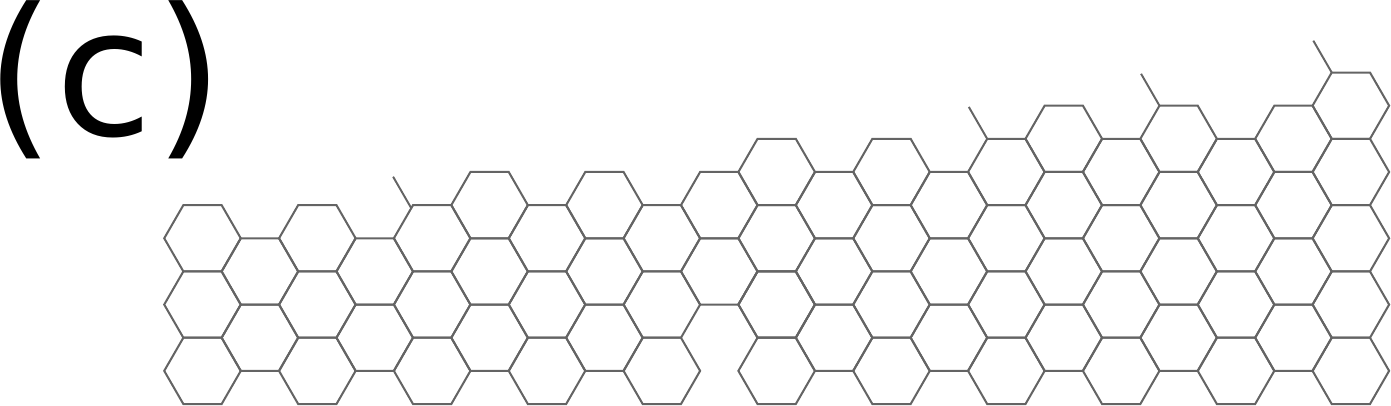} \hspace{0.32cm}
  \includegraphics[width=0.35\columnwidth]{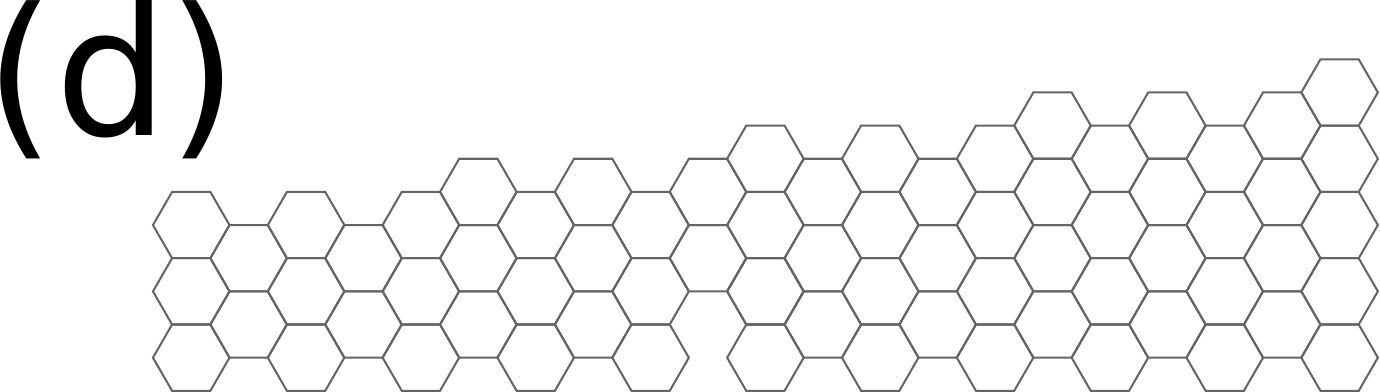} 
  \caption{ The conductance of QPCs defined in an armchair nanoribbon of width 509 atoms across the ribbon by etching (a) and bilayer patches (b). The dots mark the workpoints for the conductance mapping (see the text). (c,d) the section of the etched QPC edge with (c) or without (d) single atoms. } \label{Tef}
\end{figure}

\begin{figure}
 \includegraphics[width=0.8\columnwidth]{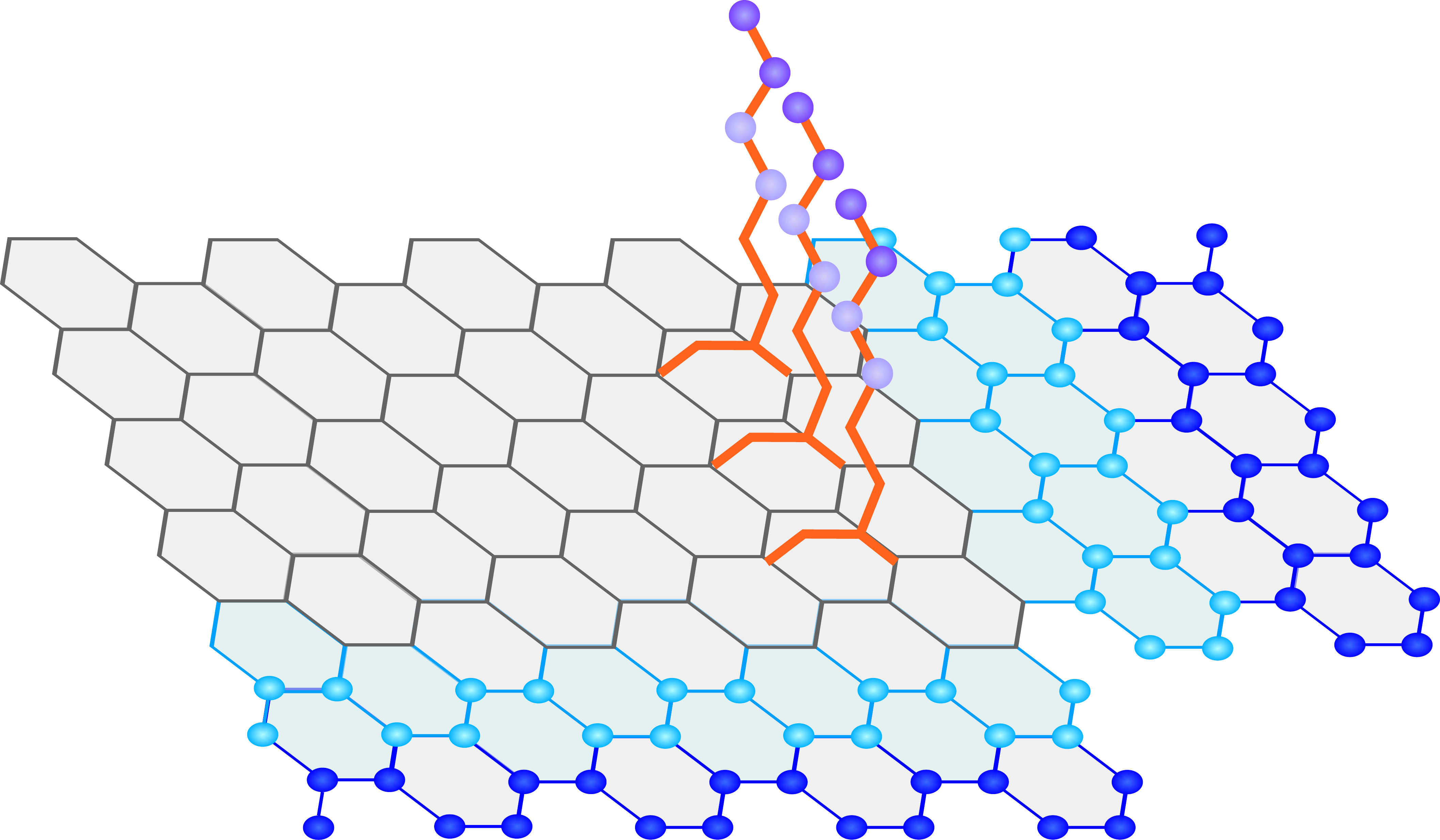}
  \caption{
The atomic structure of the corner of the computational box the blue atoms indicate the area when the horizontal leads are attached. The horizontal leads preserve the crystal structure of graphene. The atoms marked in light blue color belong to the elementary cell of the lead and the dark blue atoms form the duplicate of the elementary cell that ensures periodicity of the leads. The vertical leads are marked with the orange lines. The hopping elements are taken equal to the nearest neighbor hopping within graphene. \label{S4}
  } 
\end{figure}

 We consider an armchair nanoribbon of width $W=62$ nm,  509 atoms wide. 
 The QPC is either formed by etched out semicircles with radii $R=28$ nm
producing a  QPC [Fig.~\ref{etch_ind}(a)]
 or by bilayer patches of the same form [Fig.~\ref{etch_ind}(b)].
The QPC is $D=6$ nm wide in the narrowest point.
We consider QPC edges with a number of singly connected atoms -- similar to the ones present in the Klein edge \cite{He,KleinEdge} [Fig.~\ref{Tef}(c)] as well as ''clean'' edges  with the  singly connected atoms removed [Fig.~\ref{Tef}(d)]. 
For the SGM modeling,  open boundary conditions (BCs) at the horizontal edges at the output QPC side are applied. 
We add to the right of the QPC -- i.e. the output side -- two  leads, that are semi-infinite  in the $y$ direction and extend all along the upper and lower edge of the nanoribbon [Fig.~\ref{S4}(c)]. The extra
leads are introduced to simulate an infinite graphene sheet to eliminate the effects of the backscattering from the nanoribbon edges and the subband quantization that produces a set
of subband-dependent Fermi wavelengths instead of a single one.
Upon attachment of the leads, the corners of the computational box -- between the right lead and the top or bottom leads [Fig.~\ref{etch_ind}(c)], still act as scattering centers and produce an artificial interference.

To eliminate the scattering by the corners - which influences the SGM maps -- 
we added in the upper-right and lower-right corners two leads that are semi-infinite in the $z$-direction [Fig.~\ref{S4}] 
 that absorb the current that has not entered the in-plane leads. The additional vertical leads are attached to the corners of the computational box as the sinks of the current. As the present approach is based on the atomistic tight-binding procedure we had to choose an atomic structure form of the leads.



\begin{figure}
  \includegraphics[width=0.48\columnwidth]{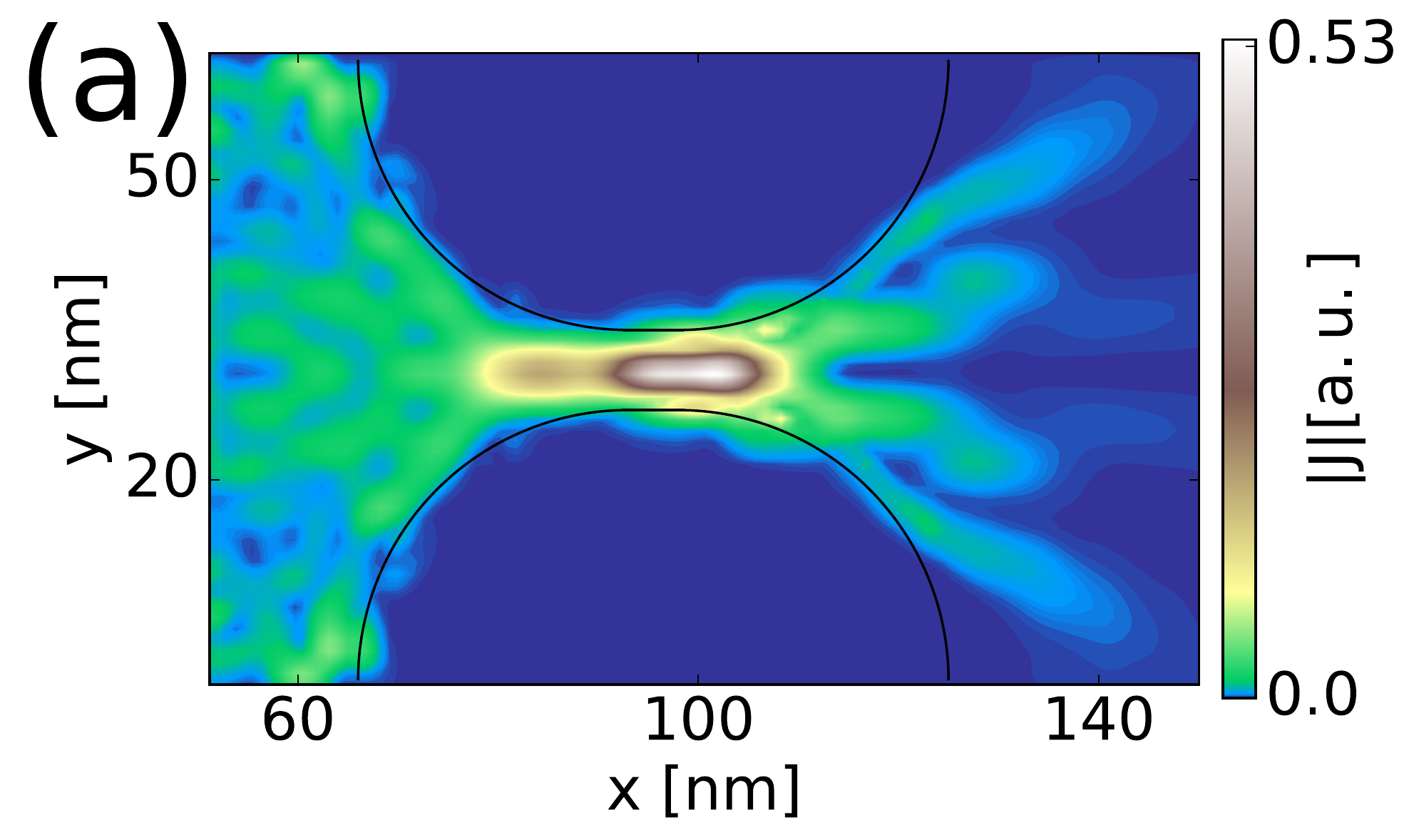}
  \includegraphics[width=0.48\columnwidth]{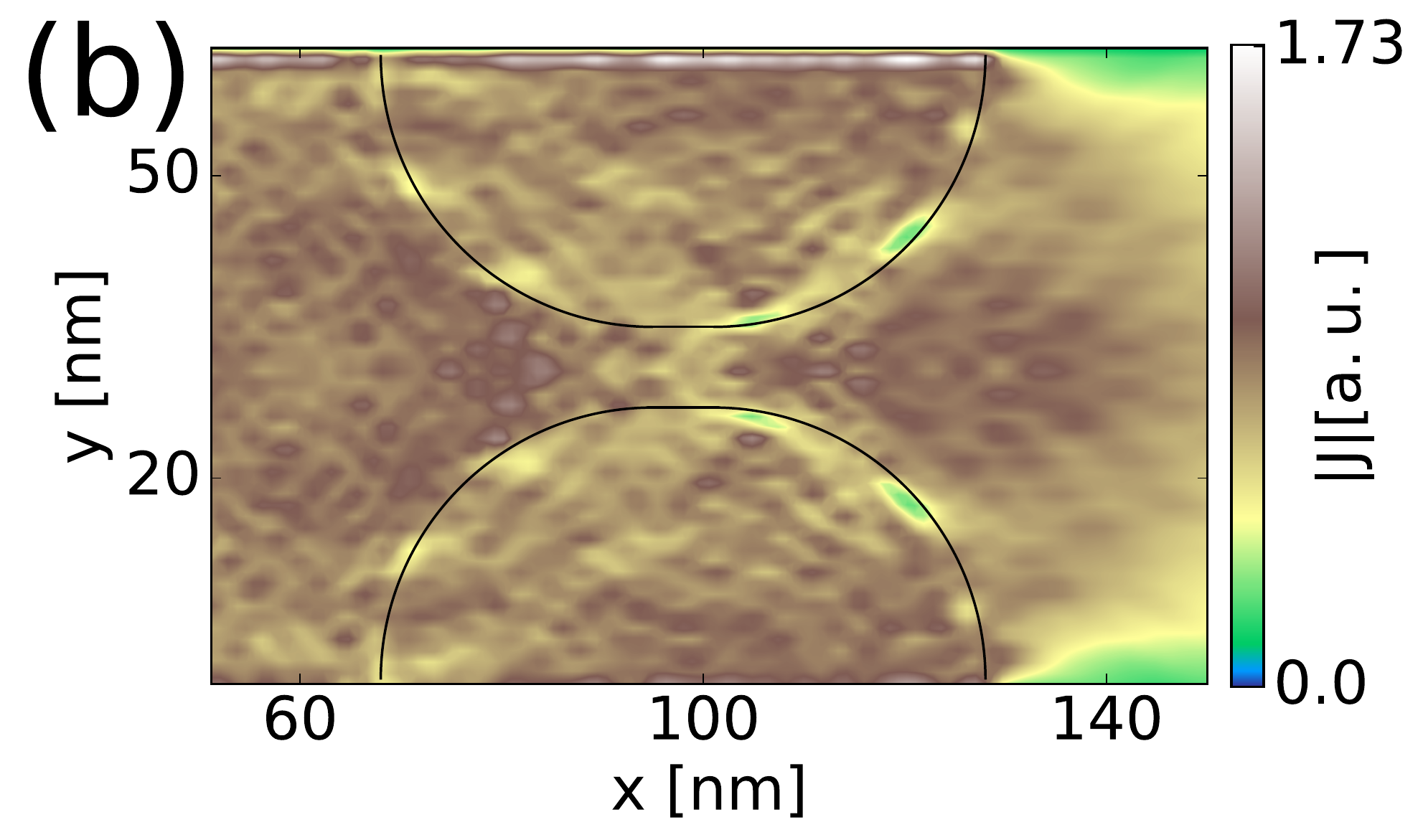}
  \caption{  The current densities in the QPC formed by biased bilayer patches at $E_F=0.327$ eV (a), within the energy gap of for bilayer patches that is (0.19,0.64) eV.  In (b) $E_F=0.764$ eV  exceeds the bias and current flows across entire ribbon.} \label{currBias}
\end{figure}

\section{Results.} 

\subsection{Conductance of the QPC constrictions} In Fig.~\ref{Tef}(a,b) the transmission probability as a function of the Fermi energy is presented.
A transport energy gap due to the constriction is present near the Dirac point.
For QPC with singly connected atoms at the etched edge [blue line in Fig.~\ref{Tef}(a)], the conductance exhibits a number of sharp peaks.
No well developed plateaus are observed, and the conductance is much lower than the one for  a uniform ribbon of the width of the narrowest part of the QPC (dashed line). This is caused by  strong backscattering  by the atomic-scale roughness of the etched QPC induced by the singly connected atoms. Upon their removal [cf. Fig.~\ref{Tef}(c) and \ref{Tef}(d)], the conductance [the orange line in Fig.~\ref{Tef}(a)] becomes a smooth function of the energy and approaches the maximal conductance for the QPC width.

For the bilayer patches we assume that the potential on the lower graphene layer is $V=0$, and $V_{b}$ on the upper layer
($V_{b}=0.64$ eV unless stated otherwise). For that bias within the finite size bilayer patches a bandgap is formed in the range of (0.19,0.64) eV.  For the Fermi energy $E_F$ of the leads within gap opened by the interlayer bias in the constriction, the current doesn't penetrate the patches [Fig.~\ref{currBias}(a)]. For $E_F$ beyond the forbidden range the current flows  across the patches [Fig.~\ref{currBias}(b)].
 Similar as the etched QPCs, the geometry of the QPC is  specified once the sample is produced, however the bilayer-patched systems can be controlled via the electric field which allows us to turn on and off the quantizing properties, or alter the number of conducting modes in the QPC.

The conductance of the patched QPC is presented in Fig.~\ref{Tef}(b) as a function of $E_F$. The dashed line shows the conductance of a uniform nanoribbon with two rectangular bilayer patches along the entire ribbon with a width matching the conditions of the narrowest point of the QPC constriction.
The conductance of the QPC with patches that contain a Klein edge and those without it, is in both cases smooth, however 
there is the ubiquitous backscattering that makes the conductance lower than that of the uniform ribbon of the same structure as the narrowest part of the QPC. With the singly connected atoms the $G(E_F)$ dependence is smoother in the patched QPC [Fig.~\ref{Tef}(b)] than for the etched one [Fig.~\ref{Tef}(a)] since even for the atoms of the upper layer that have only one neighbor in-plane, there is a non-zero hopping to the atoms in the lower layer. 


\subsection{Simulation of the scanning gate microscopy.}

For the $E_F<V_{t}=1.25$ eV, the tip introduces an n-p junction.
For  QPCs without the singly-connected atoms we choose the workpoint for the scanning maps 
at  the conductance step ($G\approx G_0$) and at the plateau $G\approx 2 G_0$. 
For the etched QPC the plateau and the step are taken at $E_F=0.312$ eV ($G=1.01 G_0$, see the orange point in Fig.~\ref{Tef}(a)) and $E_F=0.37$ eV at the etched nanoribbon ($G=1.73 G_0$, see the green point in Fig.~\ref{Tef}(a)), respectively.
For the patched QPC we take  $E_F=0.37$ meV for the plateau  ($G=1.8 G_0$,  the green point in Fig.~\ref{Tef}(b))
and $E_F=0.327$ eV for the step ($G=1.19 G_0$, the orange point in Fig.~\ref{Tef}(b)).

\begin{figure}
  \includegraphics[width=0.8\columnwidth]{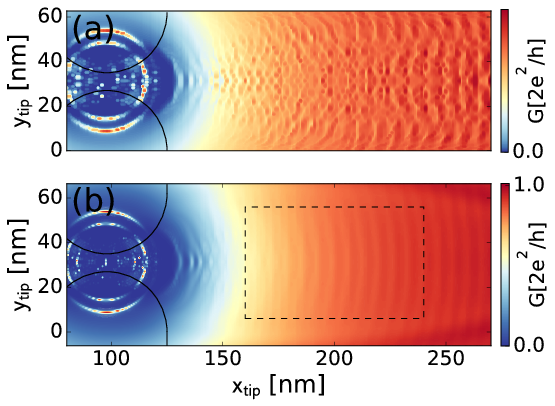}
  \caption{  
The conductance of an etched QPC without the singly connected atoms for $E_F=0.312$ eV [orange dot in Fig.~\ref{Tef}(a)] as a function of the SGM tip position for (a) 509 atom wide nanoribbon at the right QPC side (closed BCs at the vertical edges) and (b) an infinite graphene halfplane simulated with open BCs.
}\label{SGMetchedNarrow}
\end{figure}

\begin{figure}
  \includegraphics[width=0.8\columnwidth]{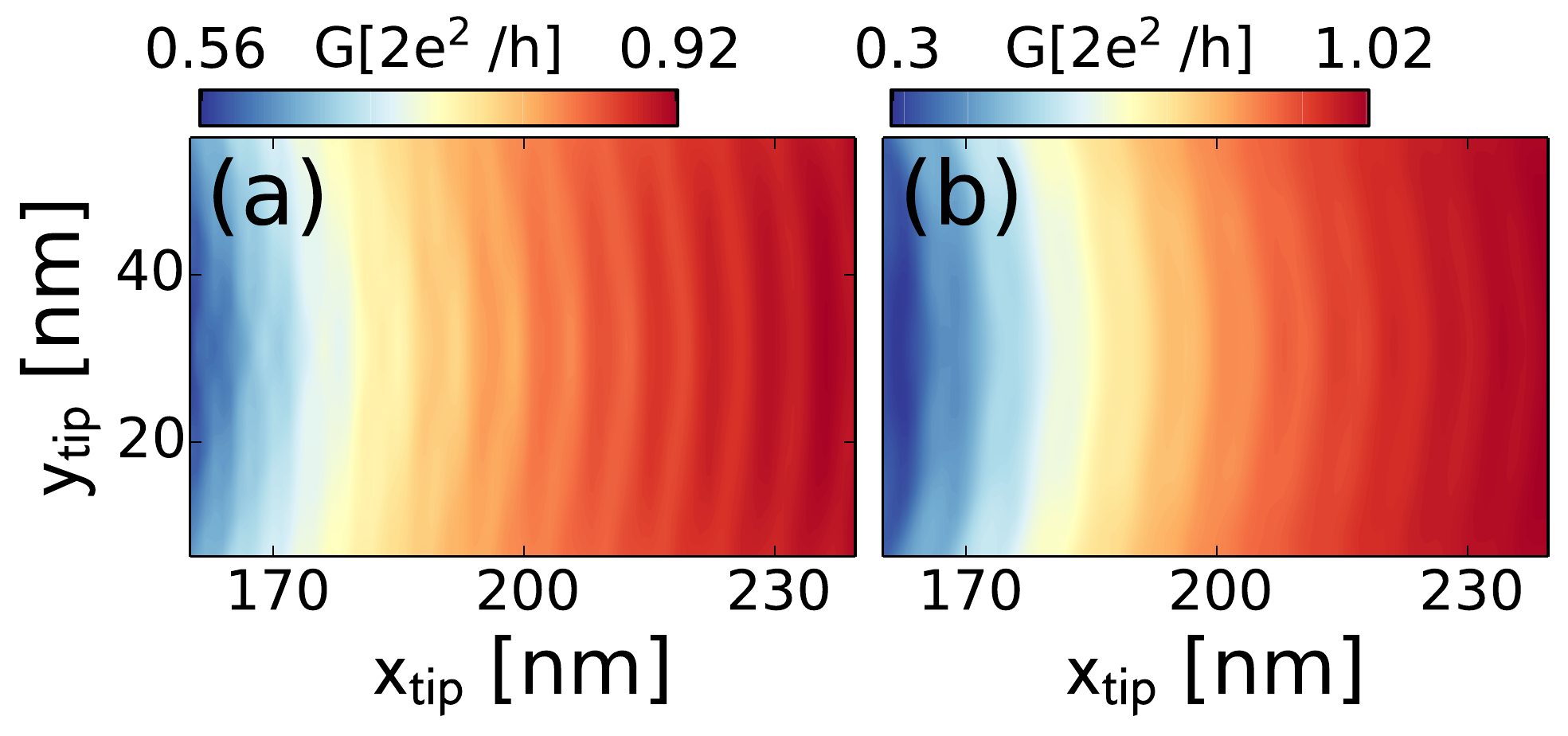}
  \caption{
Map of conductance within the region marked by the dashed rectangle
in Fig.~\ref{SGMetchedNarrow}(a) for (a) etched and (b) patched QPC.
In both cases a clean QPC (patch) edge was taken 
and a work point with large $dG/dE_F$ was assumed -- with $E_F=0.312$ eV (a) and $E_F=0.327$ eV (b) -- see the orange dots in Fig.~\ref{Tef}(a) and \ref{Tef}(b), respectively.
}\label{SGMetchedNarrowZoom}
\end{figure}





 For the QPC conductance -- in the absence of the tip -- the open BCs at the output side of the QPC  play no significant role. The conductance is nearly the same with rigid and open BCs for the vertical edges of the ribbon. This fact  results from a negligible scattering by the horizontal edges  
that could reverse the current back through the QPC to the input lead. However, the open conditions are crucial for the conductance mapping.

Let us  first consider conductance maps for closed 
BCs at the upper and lower edge of the ribbon, which are then actual ends of the sample.
Fig.~\ref{SGMetchedNarrow}(a) shows the conductance map for the etched QPC with the clean edge.  The contour of the bilayer patch is marked by black solid lines.
The two halos centered in the middle of QPC correspond to the tip-induced activation of the two resonances marked by orange arrows in Fig.~\ref{Tef}(a).
Away from constriction in Fig.~\ref{SGMetchedNarrow}(a) the conductance fluctuates in a non-regular way, due to a large number of transversal modes with different $k_F$. The nanoribbon  of the considered width 
have 19 modes at $E_F=0.312$ eV and 22 modes at  $E_F=0.37$ eV. The image contains the signal of 
superposition of waves with many different
Fermi wavelengths with the intersubband scattering. 


The  maps become simpler once open BCs are applied
to the right (output) side of the QPC to simulate an infinite graphene halfplane. 
In the conductance maps for the etched QPC with open BCs [Fig.~\ref{SGMetchedNarrow}(b)]
the QPC-centered halos remain the same as for the closed BCs [Fig.~\ref{SGMetchedNarrow}(a)].
 The difference occurs to the right of the QPC, where the simulated flake is infinite.
Far from the QPC periodic oscillations of conductance are present.
Figure \ref{SGMetchedNarrowZoom} shows the zoom at the region [dashed line in Fig.~\ref{SGMetchedNarrow}(b)] 
for the etched [Fig.~\ref{SGMetchedNarrowZoom}(a)] and the patched [Fig.~\ref{SGMetchedNarrowZoom}(b)] QPC.
In both scans the oscillations differ by an offset and not by the oscillations period.

\subsection{QPC work point vs the conductance maps}
\begin{figure}
 \includegraphics[width=0.8\columnwidth]{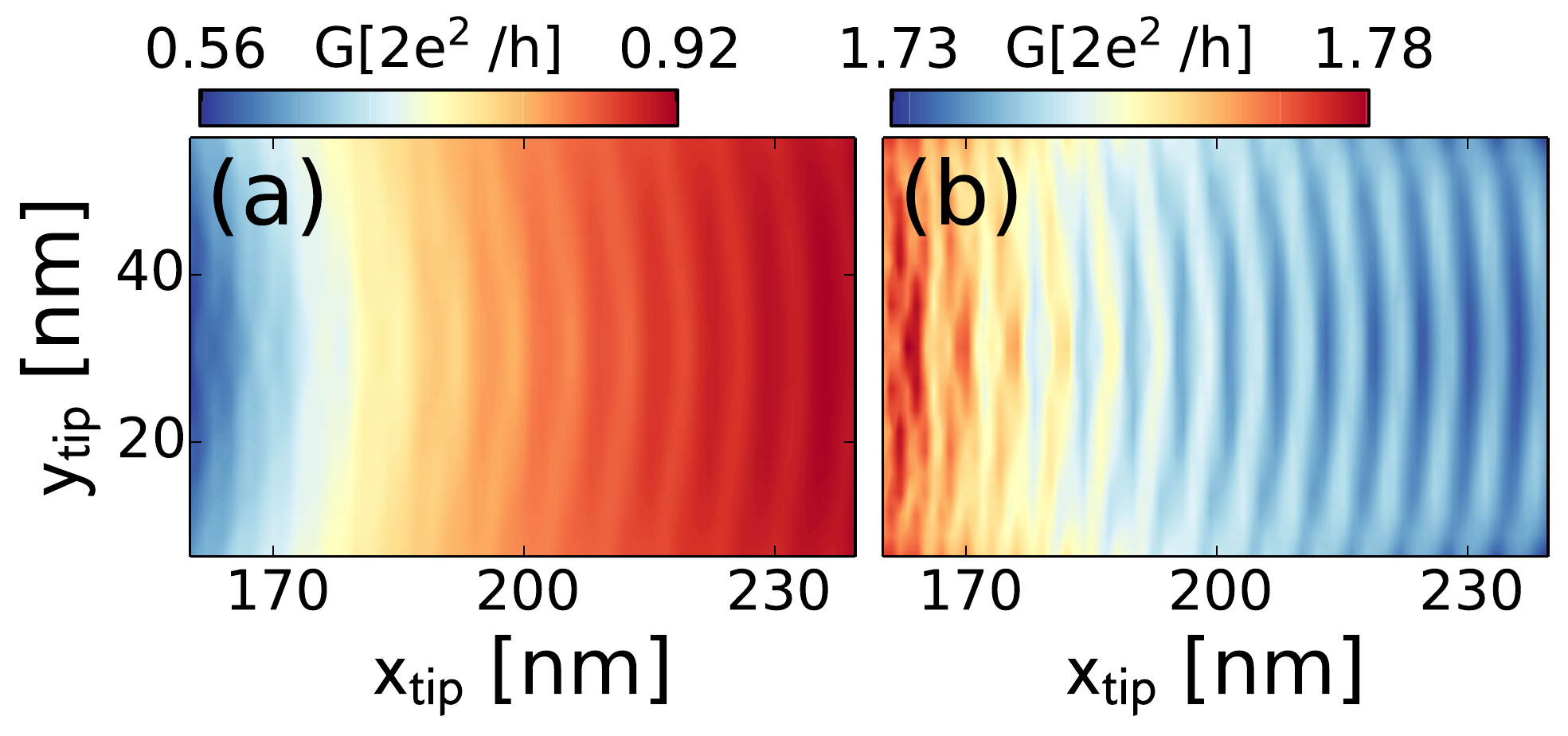}
  \caption{
The conductance maps as functions of the tip position for the etched QPCs without the singly connected atoms at the constriction. The area covered  by the map is shown in Fig.~\ref{SGMetchedNarrow}(b) by the dashed rectangle. Open boundary conditions are applied. In (a) the Fermi energy is 0.312 eV  and  in (b) 0.37 eV (b), i.e. the workpoints marked with the orange and light green points on the red conductance line in Fig.~\ref{Tef}(a). 
  } \label{s1}
\end{figure}

\begin{figure}
 \includegraphics[width=0.8\columnwidth]{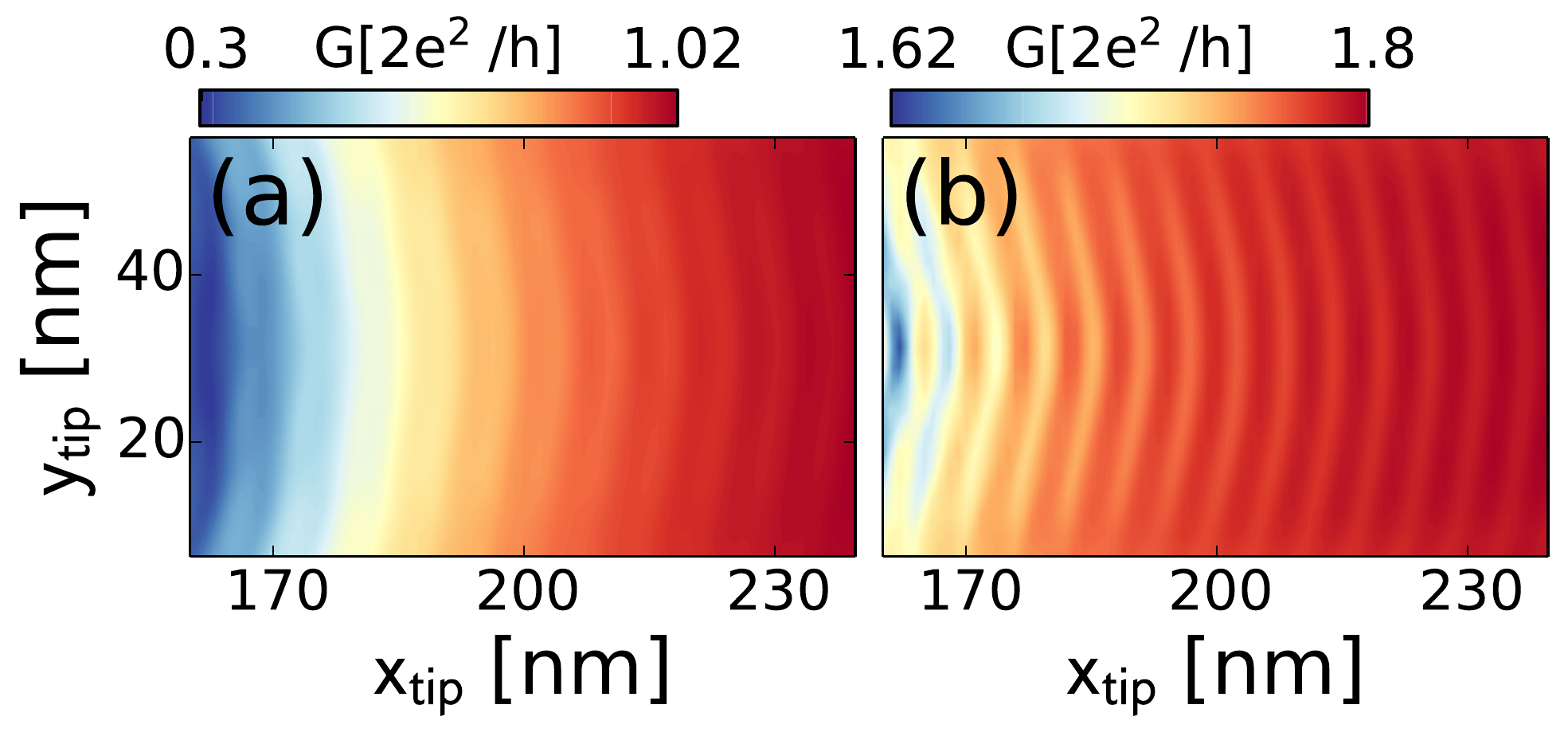}
  \caption{
Same as Fig.~\ref{s1} only for the patched QPC. The Fermi energy is 0.327 eV (a) and 0.37 eV (b) -- the workpoints are marked with the orange and light green points in Fig.~\ref{Tef}(b). 
  }  \label{s2}
\end{figure}

 The contrast of the conductance maps for fixed parameters of the tip potential depends on the Fermi energy. The contrast grows with the absolute value of $\partial G/\partial E_F$  derivative. Fig.~\ref{s1} and \ref{s2}  present the conductance maps for open boundary conditions with the workpoints marked by the orange and light green dots on the conductance vs the Fermi energy plot in Fig.~\ref{Tef}(a) and Fig.~\ref{Tef}(b) of the main text, respectively.  The maximal conductance value on the maps of Fig.~\ref{s1} and Fig.~\ref{s2} is given by the G values marked by the points on Fig.~\ref{Tef}(a) and Fig.~\ref{Tef}(b). The variation of the map increases with $G(E_F)$ slope.  The oscillation period in panels (a) and (b) of Fig.~\ref{s1} and Fig.~\ref{s2} changes in consistence with the value of the Fermi wavelength.  

\subsection{Conductance maps and edge scatterers within the constriction}

\begin{figure}
 \includegraphics[width=0.8\columnwidth]{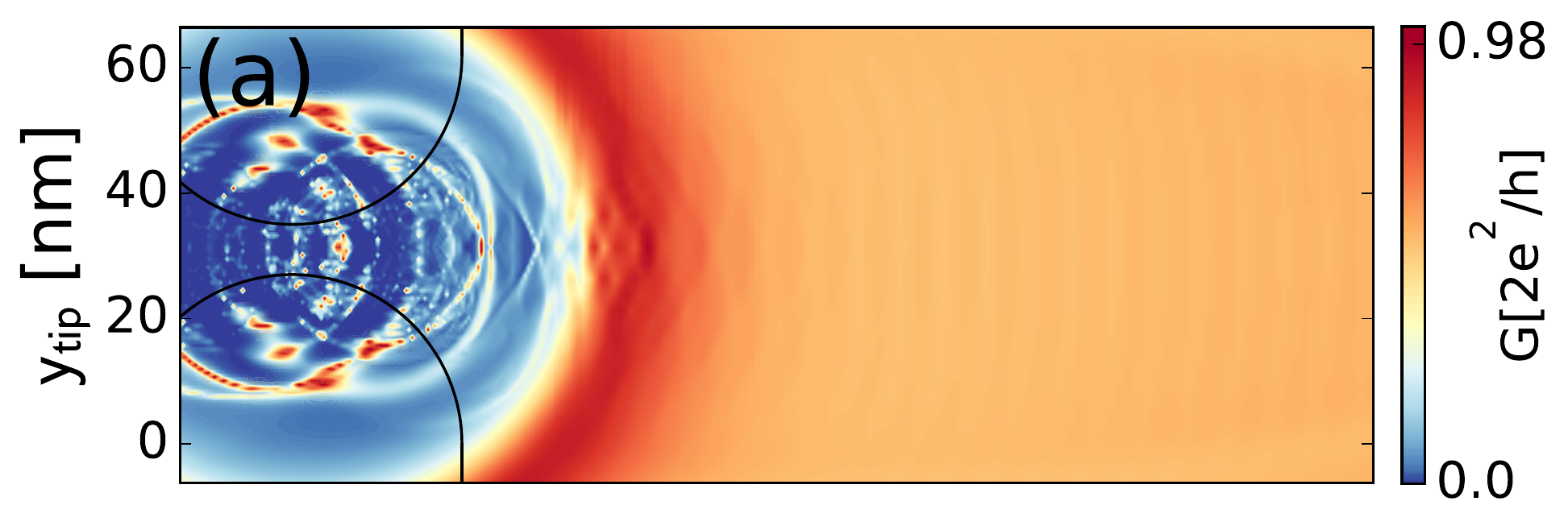}
  \caption{
The conductance as a function of the SGM tip position for nanoribbons with etched QPCs, for Fermi energy 0.312 eV (a) with singly connected atoms, with open boundary conditions and vertical probe.  \label{s3}
  } 
\end{figure}

	The singly-connected atoms within the constrictions [Fig.~\ref{Tef}(c-d)] are strong source of  scattering for electron waves that cross the QPC. The conductance is decreased when they are present within the constriction [cf. the blue and red lines in Fig.~\ref{Tef}]. The conductance maps for the etched QPC are given in Figures \ref{SGMetchedNarrow}(b) and \ref{s3}. The map changes within the constriction, but the oscillation period that is due to the backscattering at a distance from the QPC, that the paper is about, remains unchanged.


\subsection{Conductance maps with bulk scatterers off the constriction }

\begin{figure}
 \includegraphics[width=0.75\columnwidth]{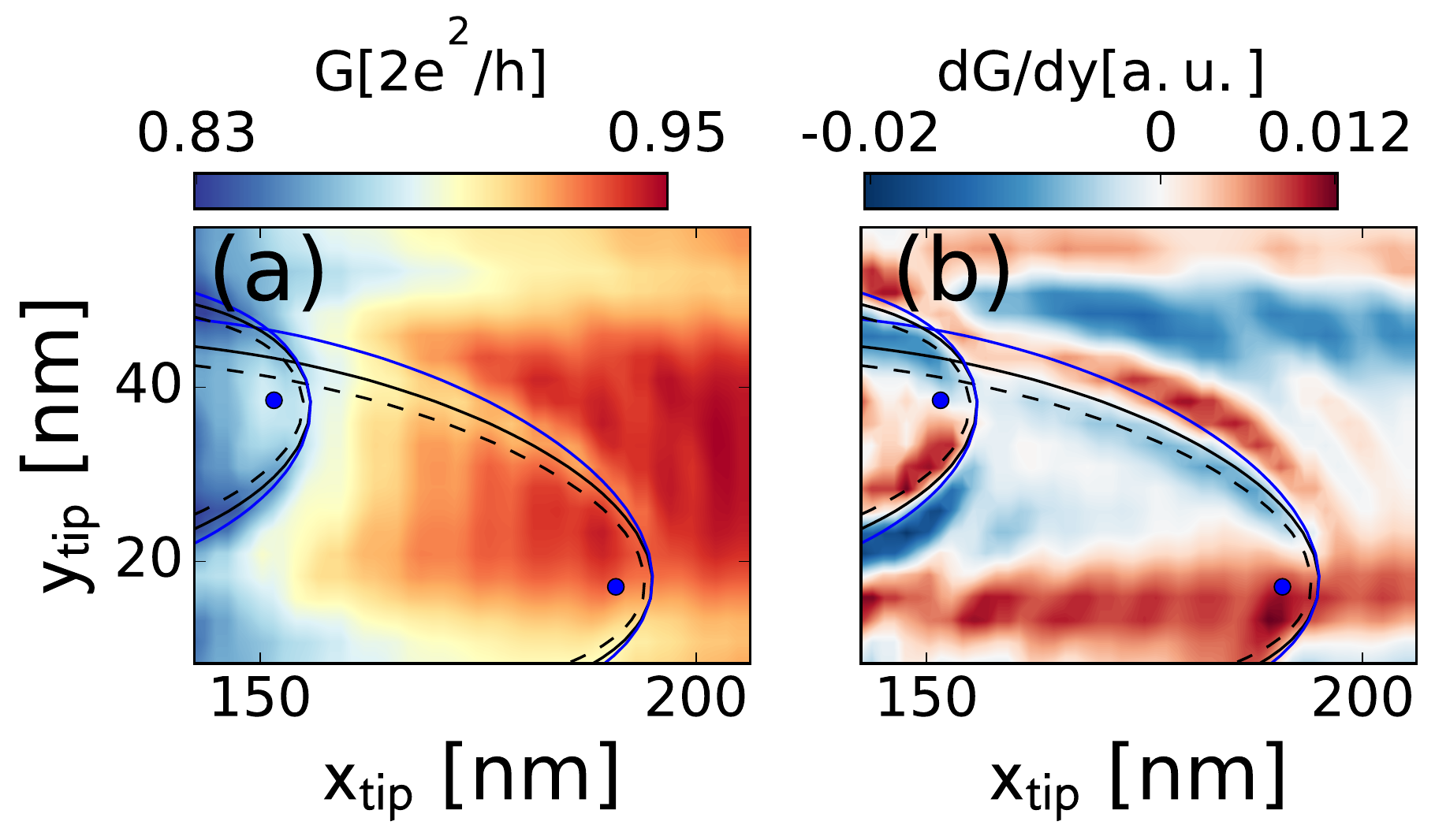}
  \caption{
The conductance as a function of the SGM tip position for nanoribbons with etched QPCs, for Fermi energy 0.37 eV (a) with fluorine adatoms and (b) derivative with respect to vertical axis of (a) to enhance the visibility of the ellipse-like fringes. Fluorine adatoms positions are marked by blue circles.  \label{fluorMaps}
  } 
\end{figure}

We consider the influence of scattering by strong local perturbation due to the fluorine adatoms bound to the carbon lattice
at a distance from  the constriction.  Two adatoms are considered with the positions marked by dots in Fig. \ref{fluorMaps}.
 We perform a modeling of scanning gate microscopy of fluorinated graphene with $V_t=0.5$ eV.
In the conductance map we observe elliptical features near the adaomts superimposed on the conductacnce oscillation pattern with the period of half the Fermi wavelength
characteristic to the clean sample. 
To improve the visibility of the signal, in Fig.~\ref{fluorMaps}(b) we plot a derivative of the conductance in Fig.~\ref{fluorMaps}(a). 
The ellipses plotted in Fig.~\ref{fluorMaps} are drawn for the conditions of the interference of the wave functions incident from the QPC and backscattered by the tip and impurity, as described in the following Section.

\section{ Discussion.} 
\begin{figure}
  \includegraphics[width=0.7\columnwidth]{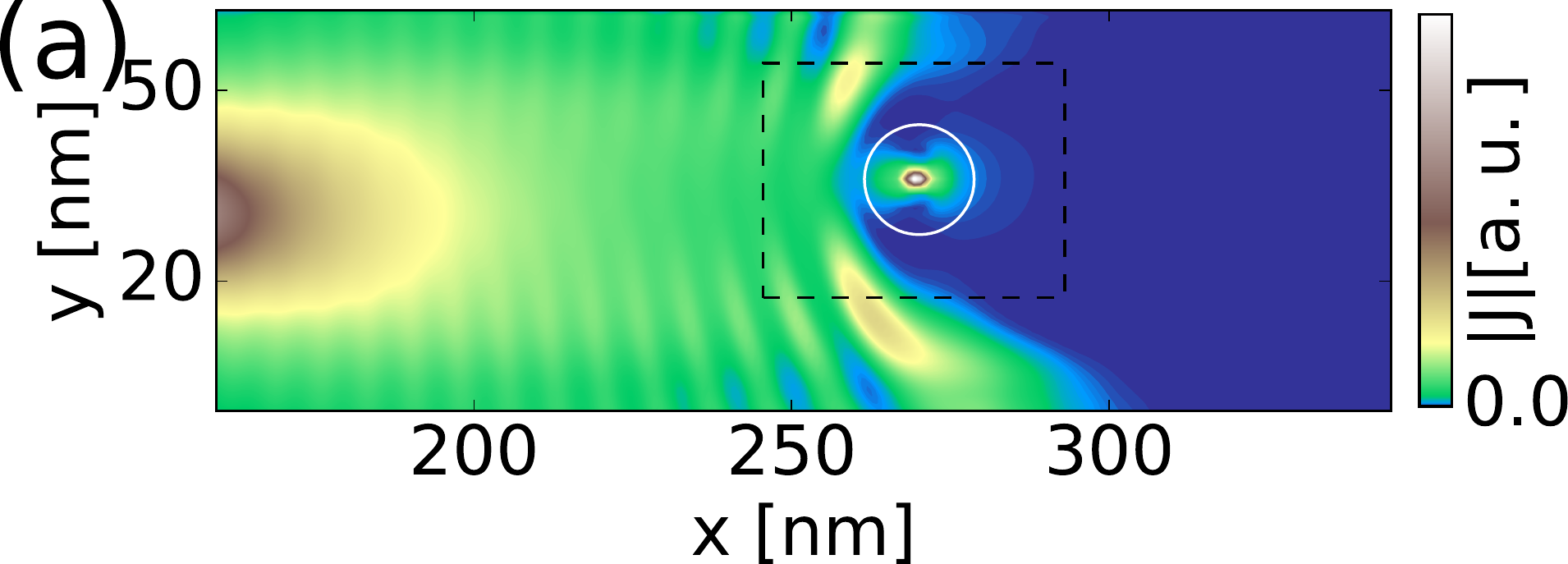}
  \includegraphics[width=0.7\columnwidth]{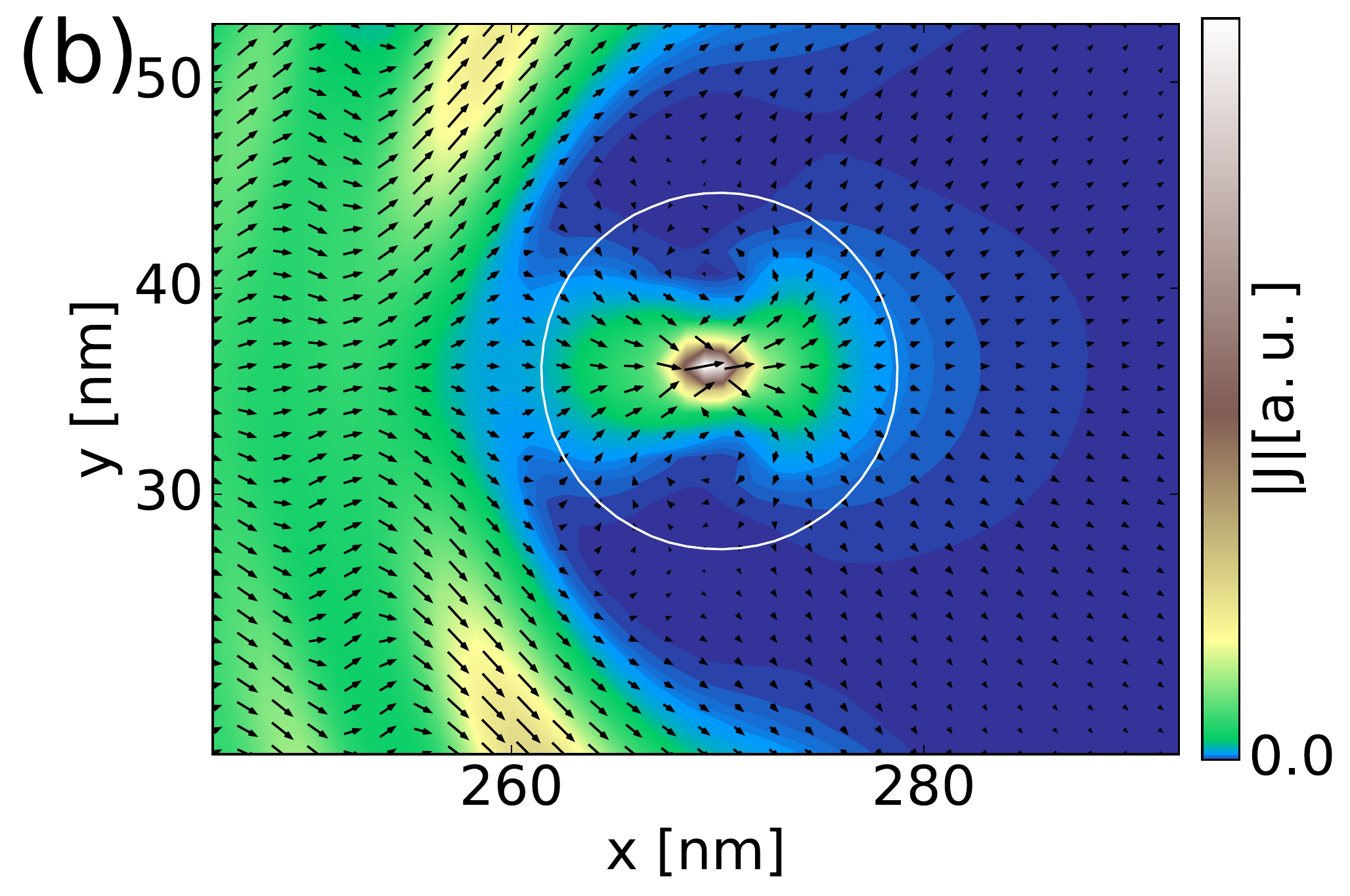}
  \caption{
(a) The current distribution for the etched QPC with the tip located at the axis
of the system for the $E_F=0.312$ eV.  The QPC center is set at $x=100$ nm. The color map shows the length of the current vector. The white circle
shows the n-p junction for $V_t=1.25$ eV.  (b) Zoom of the dashed rectangle in (a) 
 with the current orientation displayed by vectors. 
}\label{Cur}
\end{figure}

The current distribution for the etched QPC is displayed in Fig.~\ref{Cur} with the interference fringe pattern between the QPC and the tip that results from the tip-induced backscattering.
The white circle in Fig.~\ref{Cur} indicates the position where the effective tip potential  equals  $E_F$, i.e. the  n-p junction.
 In Fig.~\ref{Cur}(b) a zoom of the rectangle marked in Fig.~\ref{Cur}(a) is
displayed with the current orientation given by the vector map. Note that the current at the cross-section of the computational box is not necessarily conserved, as it can flow to the upper and lower contact. The current in the entire, infinite system however is always conserved. The current is focused by the circular n-p junction, and disperses to the right of it in Fig.~\ref{Cur}(b).

\begin{figure}
(a)   \includegraphics[width=0.32\columnwidth]{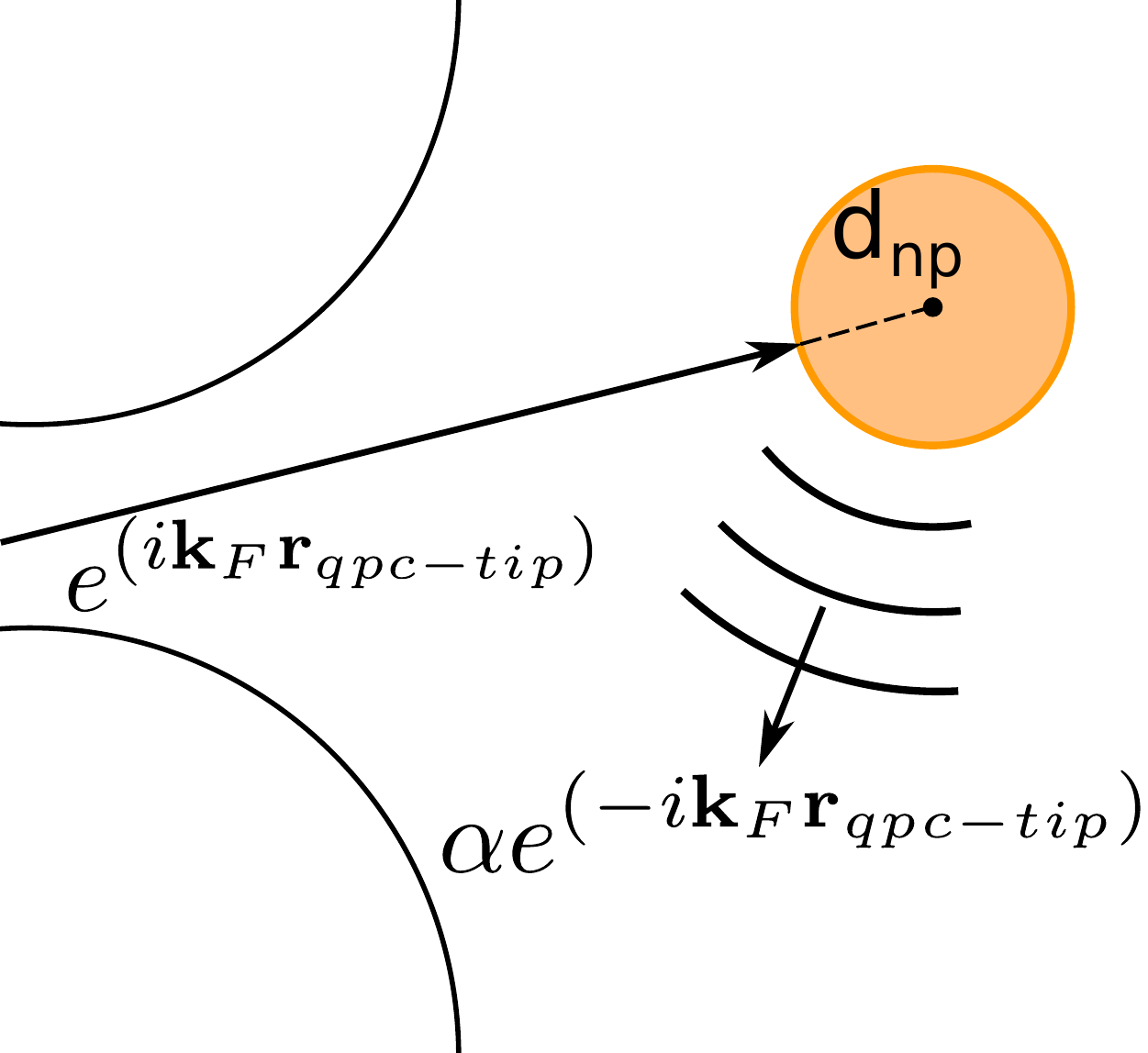}
(b)  \includegraphics[width=0.42\columnwidth]{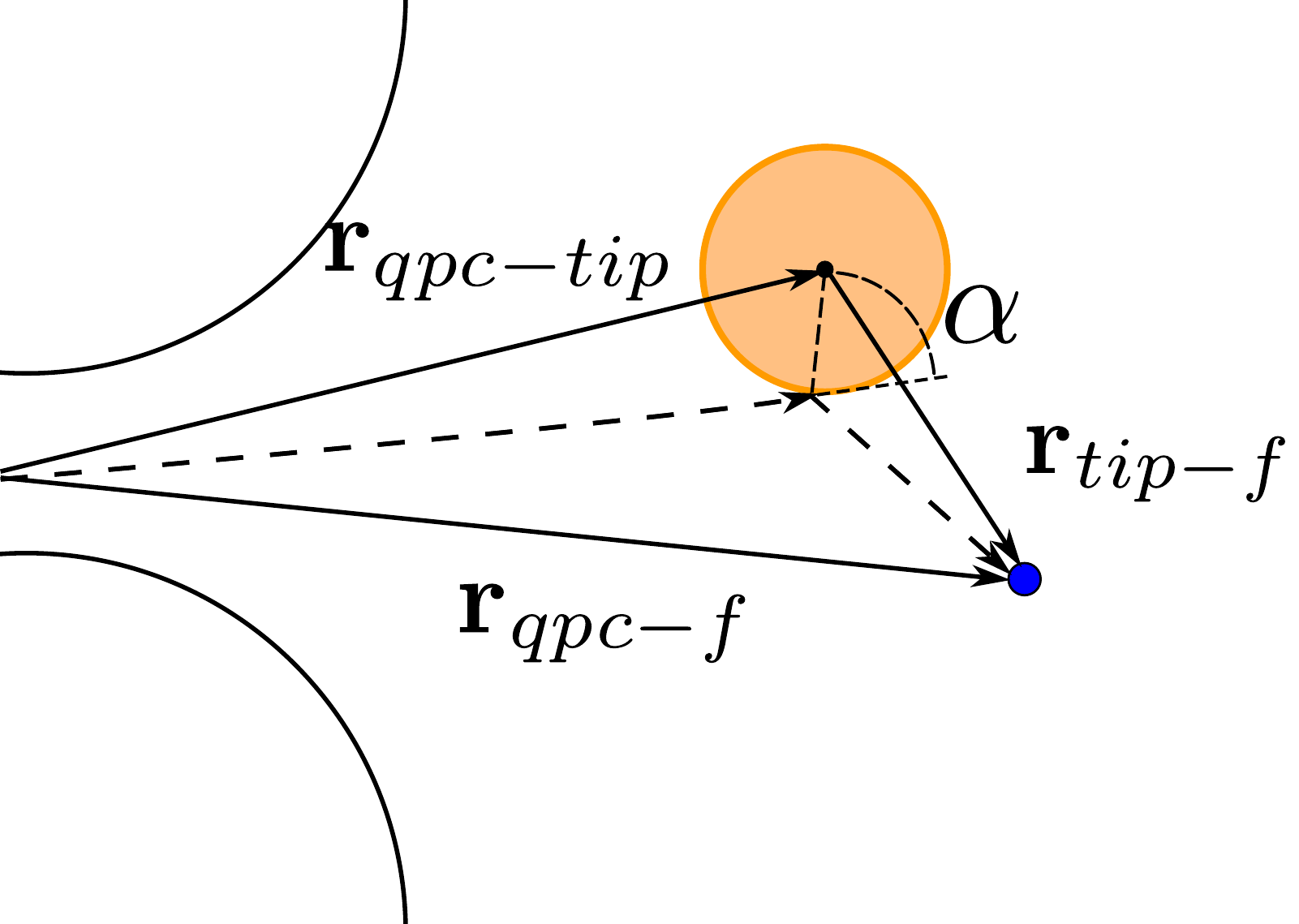}
  \caption{  
 (a) Scheme of the scattering by the tip induced n-p junction. 
 (b) Scheme of the scattering between the tip and fluorine atom. 
}\label{fig:schemScatt}
\end{figure}

\begin{figure}
  \includegraphics[width=0.65\columnwidth]{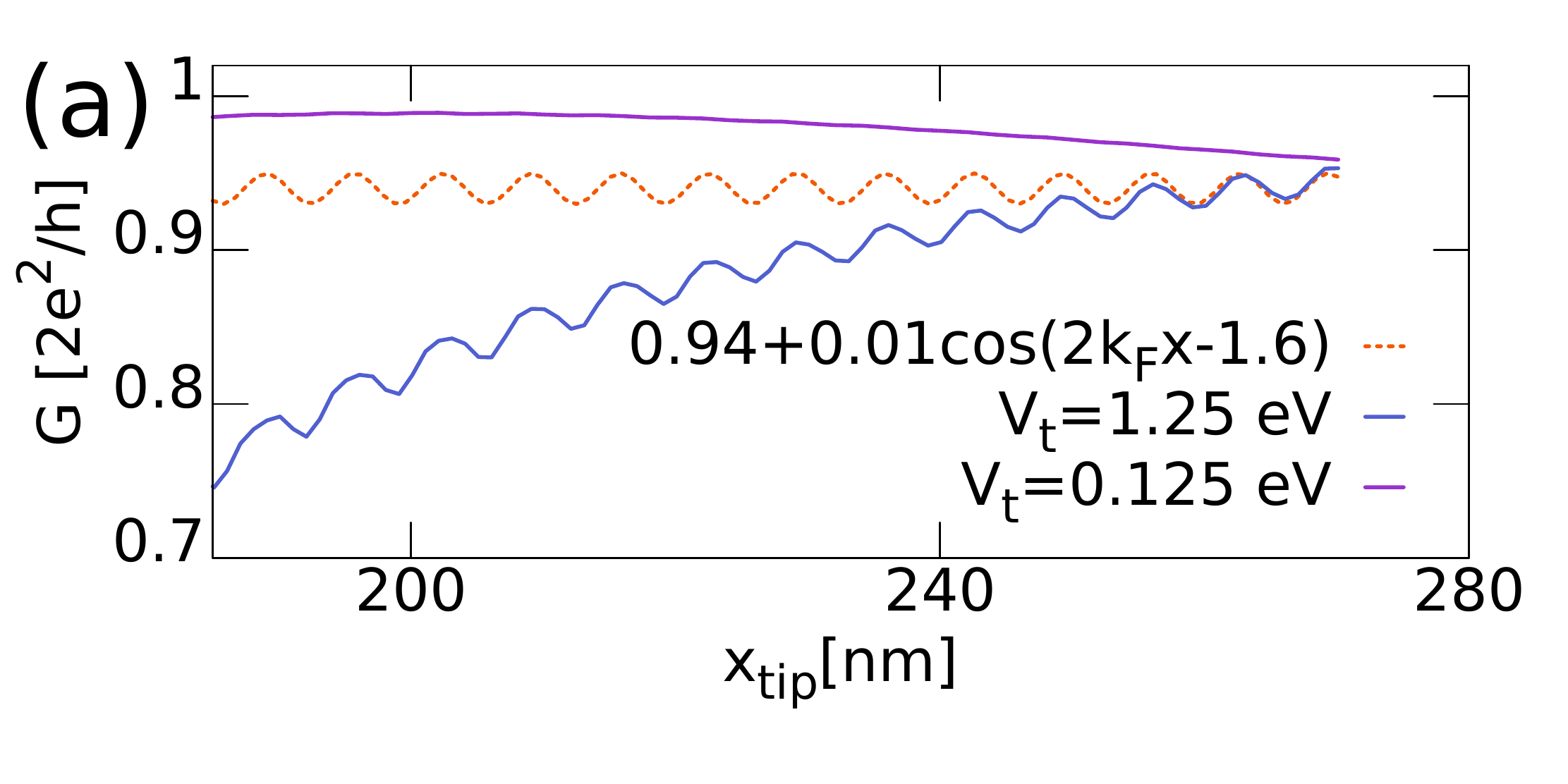}
  \includegraphics[width=0.65\columnwidth]{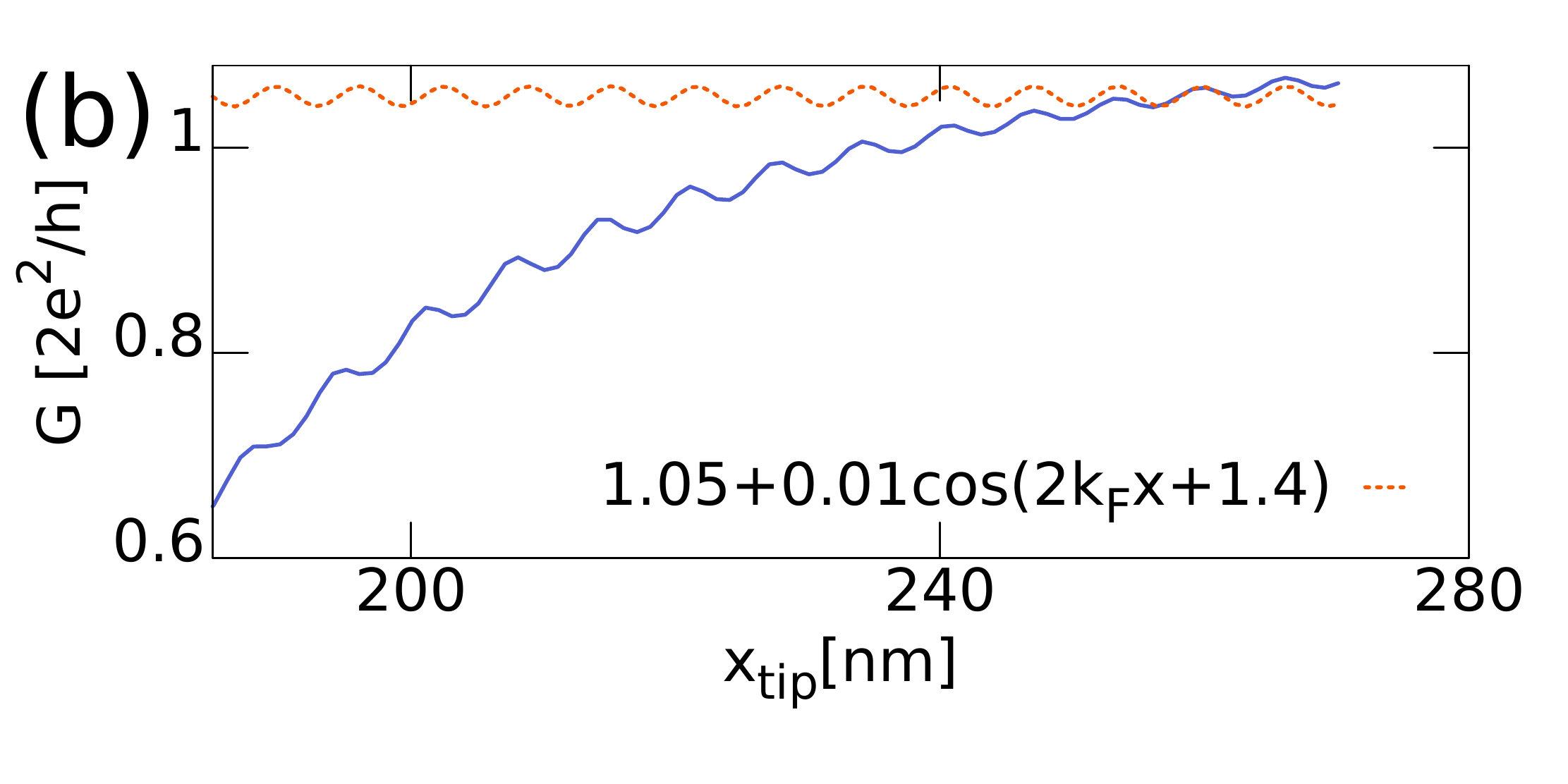}
  \caption{  
The blue lines show the cross-sections of the conductance maps along the symmetry axis of the device.
 (a) corresponds to Fig.~\ref{SGMetchedNarrowZoom}(a) for an etched QPC with $E_F=0.312$ eV, and (b) to Fig.~\ref{SGMetchedNarrowZoom}(b) for a patched QPC with $E_F=0.327$ eV.
The dashed lines indicate the cosine with $k_F$.
From Eq.~(\ref{kf}) we find $k_F=0.4695$ 1/nm for $E_F=0.312$ eV (a) and $k_F=0.493$ 1/nm for $E_F=0.327$ eV (b).
}\label{crossOscillation}
\end{figure}

In the Klein tunneling effect the Fermi electron 
incident on a perpendicular barrier larger than $E_F$ is perfectly transmitted for normal incidence angle, and the transmission probability is less than 1 for other incidence angles
\cite{Katsnelson2006, Allain2011}.  
For a non-normal incidence, the current is partially reflected, and partially transmitted and
refracted by the n-p-n junction  \cite{Lee2015, cserti2007}.
In Fig.~\ref{Cur} a normal current along the axis of the system indeed passes across the junction.
 The tip potential deflects the currents inside the central p conductivity region, and only the precisely normal component of the current passes through undeflected. Other incidence angles contribute to backscattering.

 The angular dependence of the scattering by a circular potential in graphene has been described for an incident plane wave in Ref.~\cite{cserti2007}. 
In our case the wave function incoming from the QPC opening is not a plane wave 
but it
is closer to a circular wave, which contributes to a deviation of the incidence angles from normal. 
Moreover, the tip potential that is of an electrostatic origin is bound to possess a smooth profile.
 According to Ref.~\cite{Cayssol}, for smooth potential profile the transmission probability drops deep below 100\%  already at a low deviation of the incidence angle from normal. 

Let us consider a simple model for conductance oscillations far from the QPC. The QPC is a source of a circular wave function and the SGM tip induces backscattering as argued above and shown in Fig.~\ref{fig:schemScatt}(a). 
The wave function incident from the QPC is partially reflected back to the opening (Fig.~\ref{fig:schemScatt}(a)). The incident wave $\Psi_{in}( \mathbf{r}_{tip}) = \exp (i \mathbf{k}_F ( \mathbf{r}_{qpc}-\mathbf{r }_{tip}) )$ 
and the wave backscattered by the tip $\Psi_{sc}( \mathbf{r}_{tip})= \exp( -i \mathbf{k}_F( \mathbf{r}_{qpc}-\mathbf{r }_{tip}) )$ superpose and create a standing wave between the tip and the QPC. The electron density modulation can be described by
  \begin{equation}
\left|\Psi( \mathbf{r}_{tip} )\right|^2 \propto \cos( 2 \mathbf{k}_F (\mathbf{r}_{tip} - \mathbf{r}_{qpc} ) ).
\label{scatter}
\end{equation}
  This form of the scattering density gives rise to conduction map that oscillates with the tip position, with a period of $\lambda_F/2$, where $\lambda_F=\frac{2 \pi}{ \left| \mathbf{k}_F \right| }$. The Fermi vector can be calculated for low energy from the graphene linear dispersion relation  \cite{Neto}:
\begin{equation}
 k_F=\frac{2}{3}\frac{E_F}{t a_{CC}} . \label{kf}
\end{equation}  

  In Fig.~\ref{crossOscillation} the cross sections along the axis of the system of Fig.~\ref{SGMetchedNarrowZoom}(a) and \ref{SGMetchedNarrowZoom}(b) are shown together with a cosine shifted in phase and offset  to adjust to the conductance calculated from the quantum scattering problem. Far from the QPC, the modeled conductance is close to a cosine with the $k_F$ that agrees with the wave vector obtained from the dispersion relation of graphene. 
As seen in Fig.~\ref{SGMetchedNarrow}(b) and Figs.~\ref{SGMetchedNarrowZoom}-\ref{s3}, far from the QPC the oscillations can be described by a simple model. 
In   Fig.~\ref{crossOscillation}  with the purple line we marked the results obtained for $V_t=0.125$ eV which is below  $E_F$. In this case no backscattered interference pattern is observed. 
We  find that formation of the n-p junction by the tip is a necessary condition for observation of the interference fringes.

The signal observed in the presence of scattering near the fluorine adatoms with the elliptical features in the conductance map 
of Fig.~\ref{fluorMaps} is similar to the one identified recently in III-V semiconductors \cite{Kolacha} as due to the interference
signal induced by scattering by the tip and a fixed defect. The interference paths are schematically shown in Fig.~\ref{fig:schemScatt}(b).
Figure~\ref{fig:schemScatt}(a) illustrates backscattering by the n-p junction induced by the tip resulting in the interference
pattern with half the flux quantum discussed above. In Fig.~\ref{fig:schemScatt}(b) the electron wave incident from the QPC to the fluorine
adatom interferes with the wave that is scattered by the tip-induced n-p junction. 
 The resulting conductance pattern 
 can be approximately described by 
  \begin{equation}
G \propto \cos( k_F ( r_{qpc-tip-f} - r_{qpc-f} ) ).
\label{scatterImpurity}
\end{equation}
In Fig.~\ref{fluorMaps}  with the dashed lines we plot the isolines of $ r_{qpc-tip-f} - r_{qpc-f} = \lambda_F/2 $.
 The dashed ellipse corresponds to  point-like tip,
while the solid black line in Fig.~\ref{fluorMaps} accounts for the finite radius of the n-p junction  
$d_{np} = d \sqrt{ \frac{V_t}{E_F}-1 }$.
The black solid line in Fig.~\ref{fluorMaps} was obtained for condition $ r_{qpc-tip-f} - r_{qpc-f} - d_{np} = \lambda_F/2 $.
A still closer approximation is obtained when one accounts for the dependence of the penetration depth of the electron incidence angle \cite{Kolacha}.
For the blue solid line in Fig. 10 we considered the condition $ r_{qpc-tip-f} - r_{qpc-f} - d \sqrt{ \frac{V_t}{E_F\cos(\alpha)}-1 } = \lambda_F/2 $.
Concluding, in presence of the defects the conductance map resolve the interference involving the tip, the QPC and the defect -- similarly 
as previously described for III-V semiconductors.

\begin{figure}
  \includegraphics[width=0.49\columnwidth]{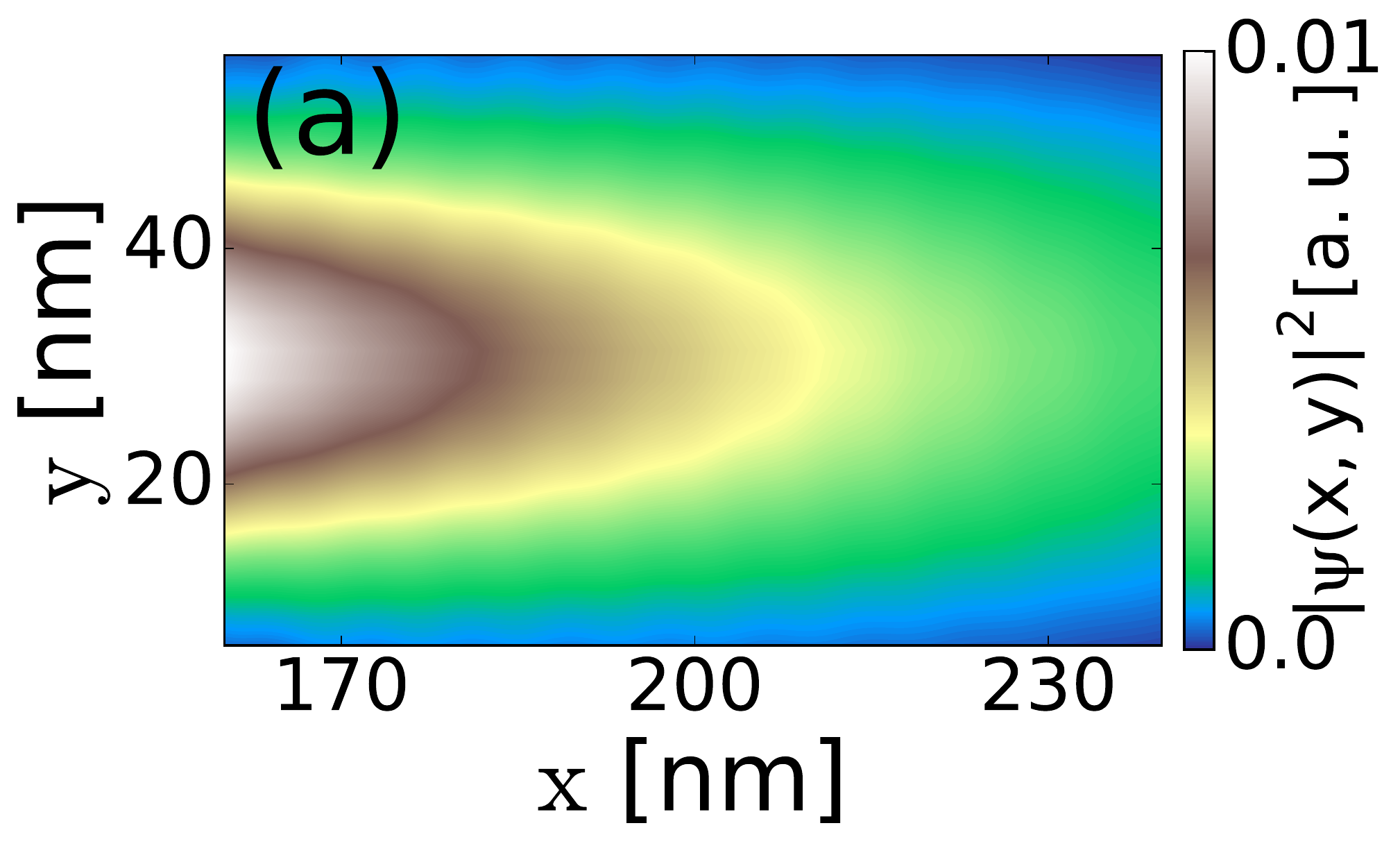}
  \includegraphics[width=0.49\columnwidth]{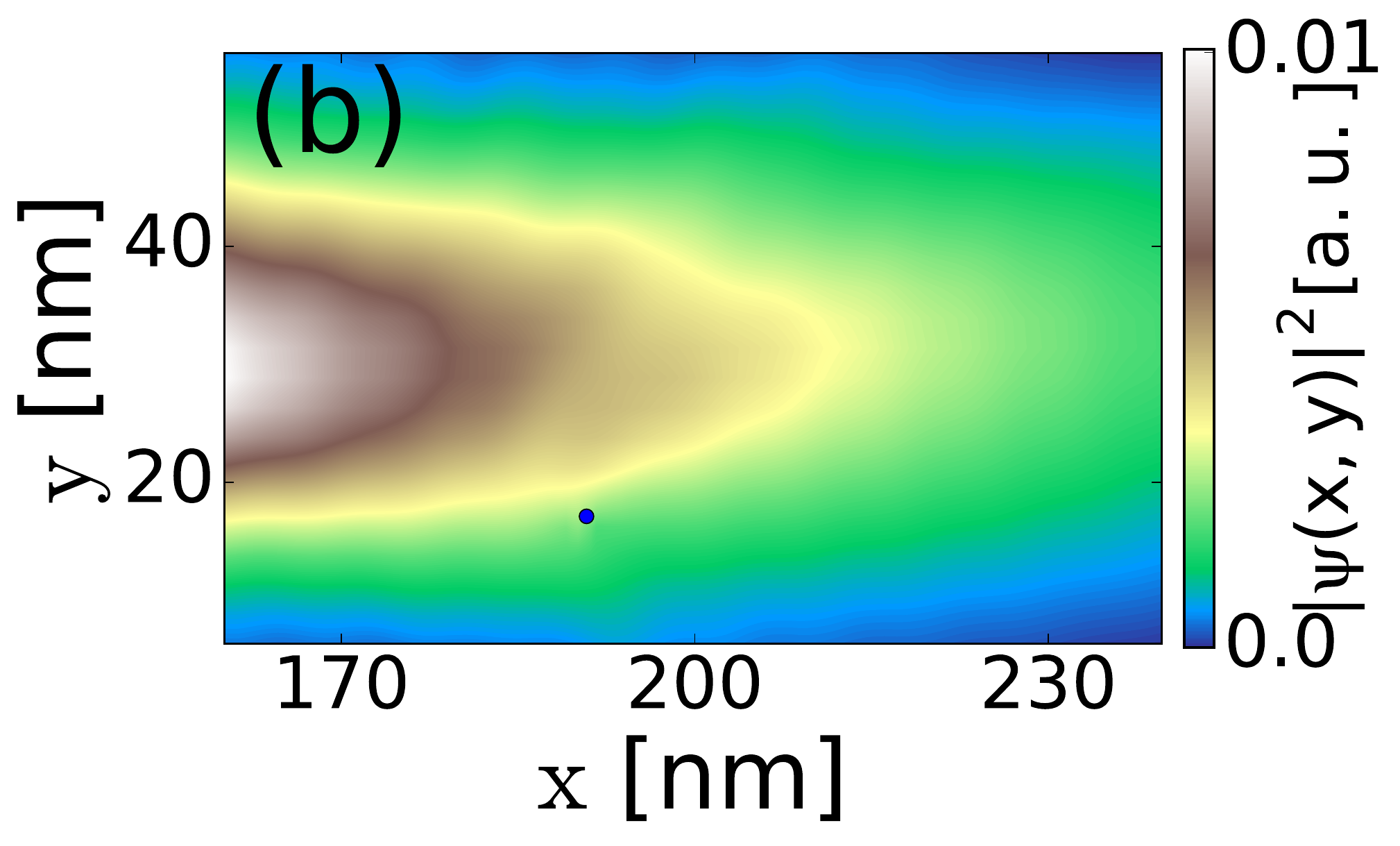}
  \caption{ The local density of states for $E_F=0.312$ eV in the absence of the tip for the etched constriction in pure graphene (a) and with fluorine adatoms (b).
}\label{fig:dens}
\end{figure}

The incomplete transmission in the Klein effect for electron incidence deviating from normal was used for construction of the n-p-n Fabry-P\'erot interferometers \cite{shytov,young} in graphene. At n-p-n junctions \cite{cheianov,xing,moghaddam,libisch}  interference of refracted waves in ballistic graphene appears in the scattering electron density that is referred to as the  local density of states \cite{cheianov}. 
The present work deals with SGM with spatial resolution of the standing waves in conductance maps and not in the local density of states only. The present idea does not require sharp n-p junctions  or a point-like injection and detection of the current as in the Veselago lensing \cite{cheianov,xing,moghaddam,libisch}. 
Fig.~\ref{fig:dens} shows the probability density without the SGM tip for the electrons incoming from the left terminal for pristine graphene (a) and with fluorine adatoms (b). 
For the dilute fluorinated graphene the backscattering by the fluorine adatom is resolved in the density plot in Fig.~\ref{fig:dens}(b). There is no correlation between the densities and conductance maps.
The SGM maps resolve the quantum transport properties or in-plane conductance of the sample when the tip becomes the source of an additional scattering.
In contrast to the 2DEG in III-V quantum wells the surface electron gas in graphene can be alternatively studied with the scanning tunneling microscopy (STM) \cite{cheianov,xing,moghaddam}.
In this configuration the tunneling microscope acts as a contact and not as a gate electrode. Instead of the scattering effects involving the tip 
the STM resolves the local density of states.

Ref.~\onlinecite{Neubeck} provided a SGM map of a graphene QPC for  nominal tip potential set $V_t=-0.5$ eV. The resistance map  of this work  \cite{Neubeck}
resolved only the QPC itself and not the interference fringes that were described here.
The nominal $V_t$ value given in Ref.~\onlinecite{Neubeck} is an unscreened parameter,
and it is not granted that the screened tip potential was strong enough to induce formation of the n-p junction, since no control of $E_F$ was demonstrated \cite{Neubeck}.
Nevertheless, the present work indicates that
observation of the spatial maps of the backscattering interference pattern in graphene is not excluded by the Klein tunelling effect. 

\section{Summary and Conclusions}

We have studied the current constriction by graphene QPCs formed by a gap between two biased bilayer
 patches and by a narrowing of a graphene ribbon using an atomistic tight binding method and a Landauer approach.
We considered conductance mapping as a function of a floating probe position. For this purpose open boundary conditions on the output side of the QPC
were introduced in order to produce an image clean from  backscattering by the edges and consequences of multiple Fermi wavelengths resulting from subband quantization. 
With the open boundary conditions simulating an infinite graphene plane at the output side of the QPC we found a clear interference pattern in the conductance map with the period of half the Fermi wavelength characteristic to the backscattering by the tip. The interference is only observed
provided that the tip induces an n-p junction in graphene. The backscattering of the electron wave function that occurs by the circular
tip induced n-p junction for  electron incidence at angles that deviate from normal is enough for the interference to be observed 
in the calculated conductance images. The finding that the Klein effect does not prevent observation of the standing waves induced by the tip in graphene opens perspectives for experimental determination of the current distribution, current branching by scattering defects, coherence length, etc.


%


{\it Acknowledgments}
This work was supported by the National Science Centre (NCN) according to decision DEC-2015/17/B/ST3/01161.
The calculations were performed on PL-Grid Infrastructure.

\bibliography{RingBib}

\end{document}